\newcommand{\ra}{\rightarrow}
 \newcommand{\Om}{\Omega} \newcommand{\pl}{\partial}
 \newcommand{\bq}{\begin{equation}}
\newcommand{\eq}{\end{equation}} \newcommand{\ph}{\varphi}
\newcommand{\iy}{\infty}
\newcommand{\ps}{\psi}  \newcommand{\dl}{\delta} \newcommand{\Ps}{\Psi}
\newcommand{\ch}{\raisebox{.4ex}{$\chi$}} \newcommand{\ov}{\over}
\newcommand{\ve}{\varepsilon} \newcommand{\De}{\Delta}
\newcommand{\be}{\beta} \newcommand{\al}{\alpha} \newcommand{\hf}{{1\ov2}}
\newcommand{\noi}{\noindent}
\newcommand{\Es}{{\cal E}_s} \renewcommand{\bar}{\overline}
\newcommand{\tl}{\tilde} \newcommand{\bR}{{\bf R}}
\newcommand{\re}{{\rm Re\;}} \newcommand{\im}{{\rm Im\;}}
\newcommand{\bC}{{\bf C}}\newcommand{\ua}{\uparrow}\newcommand{\da}{\downarrow}
\newcommand{\cH}{{\cal H}} \newcommand{\Ga}{\Gamma} \newcommand{\Ph}{\Phi}
\newcommand{\st}{\raisebox{.0ex}{*}} \renewcommand{\sp}{\vspace{.1in}}
\newcommand{\cI}{{\cal I}} \newcommand{\norm}{\parallel}

\documentstyle[12pt]{article} \textwidth=6in \oddsidemargin=.3in
\begin{document}

\begin{center}{\large\bf Asymptotics for the Fredholm Determinant of the Sine
Kernel
on a Union of Intervals} \end{center}\sp

\begin{center}{{\bf Harold Widom}\footnote{Supported by National Science
Foundation
grant DMS-9216203.}\\
{\it Department of Mathematics\\
University of California\\
Santa Cruz, CA 95064, USA}}\end{center}
\vspace{.1in}
\begin{center}{\bf I. Introduction} \end{center}\sp

The sine kernel
\[K(x,y):={\sin (x-y)\ov\pi(x-y)}\]
arises in many areas of mathematics and mathematical physics.  There is an
extensive
literature on the asymptotics of the eigenvalues of $K_s$, the operator
with this kernel on an interval of length $s$, as $s\ra\iy$, for example
\cite{dcm,f,lw,sl}, and asymptotic fomulas of various kinds were obtained. Some
of
these derivations were rigorous, others were more  heuristic.

The Fredholm determinant of the kernel is of particular interest. In the bulk
scaling
limit of the Gaussian Unitary Ensemble of Hermitian matrices, the
probability that an interval of length $s$ contains no eigenvalues is equal to
\[\det\,(I-K_s).\]
The asymptotics of Fredholm determinants of convolution kernels $k(x-y)$ have a
long
history. (The history of their discrete analogue, Toeplitz determinants, is
even
longer, beginning with the 1915 paper \cite{sz} of G. Szeg\"o.) If the Fourier
transform $\hat{k}$ of $k$ is smooth and less than 1, and if $k$ satisfies some
other
conditions, then for the corresponding operator $K_s$ one has as $s\ra\iy$
\[\log\,\det\,(I-K_s)=c_1\,s+c_2+o(1),\]
where $c_1$ and $c_2$ are explicitly determined constants \cite{k}. If
$\hat{k}<1$ and
is smooth except for jump discontinuities then the result becomes \cite{bw}
\[\log\,\det\,(I-K_s)=c_1\,s+c_2\,\log s+c_3+o(1).\]
Even for some cases where $\hat{k}<1$ is violated at finitely many points a
similar
relation holds \cite{b}. But for the sine kernel $\hat{k}$ is the
characteristic
function of the interval $[-1,\,1]$ and so the condition is violated on a set
of
positive measure and the situation is considerably more complicated.

The asymptotics of the determinant for the sine kernel were probably first
investigated by Dyson \cite{d}. He derived the asymptotic formula
\bq\det\,(I-K_s)\sim
2^{1/3}\,e^{3\zeta'(-1)}\,s^{-1/4}\,e^{-s^2/8}\label{dy}\eq
by applying a scaling argument to a known asymptotic formula for
Toeplitz determinants \cite{w2}, and then used inverse scattering techniques to
complete this to an asymptotic expansion. In 1980 Jimbo et al.
\cite{jmms} showed that the logarithmic derivative of this determinant
satisfies a second-order differential equation which is reducible to a
Painlev\'e V
equation. The asymptotic expansion can also be obtained, starting from
(\ref{dy}),
by substituting a formal expansion into the equation and successively solving
for the
coefficients.

Another quantity of interest from the random matrix point of view is
\[\mbox{tr}\;K_s\,(I-K_s)^{-1},\]
which in the same model equals the probability that an interval of length $s$
contains
precisely one eigenvalue, divided by the probability that it contains none.
(There are
similar but more complicated formulas for any finite number of eigenvalues
\cite{m}.) The asymptotics of this were derived in \cite{btw} where
\bq\mbox{tr}\;K_s\,(I-K_s)^{-1}\sim{e^s\ov2\sqrt{2\pi s}},\label{tr}\eq
was obtained by scaling an anologous result for
Toeplitz matrices, which in turn was proved by exploiting its connection with
orthogonal polynomials, whose asymptotics were known. This formula was then
extended to
a complete asymptotic expansion by using the differential equation.

None of these asymptotic results were rigorously proved by the methods
described. In
\cite{w3}, by studying the asymptotics of a continuous analogue of orthogonal
polynomials, we were able to give a proof of (\ref{tr}) and of the first-order
asymptotics in (\ref{dy}),
\[\log\,\det\,(I-K_s)\sim-s^2/8,\]
actually the slightly stronger
\bq{d\ov ds}\log\,\det\,(I-K_s)=-s/4+O(1).\label{det}\eq
As far as we know nothing beyond this has actually been proved up to now, and
refinements
of these are not the subject of this paper. The subject is the extension of
this
work from the one interval case to several intervals. Thus a single
interval of length $s$ is replaced by $sJ$, where $J$ is a fixed union of
intervals.
Even these asymptotics are quite elaborate and involve hyperelliptic integrals
and the
Jacobi inversion problem for Abelian integrals.

We continue to denote our operator by $K_s$, and it is understood that it now
acts on $sJ$. We shall show that as $s\ra\iy$
\bq-{d\ov ds}\log\,\det\,(I-K_s)=c_1\,s+c_2(s)+o(1),\label{Det}\eq
where $c_1$ is a constant and $c_2(s)$ is a certain bounded oscillatory
function of
$s$, and that
\bq\mbox{tr}\;K_s\,(I-K_s)^{-1}={e^{c_3\,s}\ov\sqrt s}(c_4(s)+O(s^{-\hf})),
\label{Tr}\eq
where $c_3$ is another constant and $c_4(s)$ is another bounded
oscillatory function. The constants $c_1$ and $c_3$ are explicitly
computable but the determination of the functions
$c_2(s)$ and  $c_4(s)$ requires the solution of a Jacobi inversion problem. (At
least
this is the case with this method. It is possible that a different approach
might lead to simpler representations.) If $J$ consistes
of $m$ intervals then there is a curve, parametrized by $s$, in the $m-1$-torus
${\bf T}^{m-1}$, and two real-valued functions defined on ${\bf T}^{m-1}$.
These
functions, when restricted to the curve, are $c_2(s)$ and $c_4(s)$.
When $m=2$ they are periodic, and
we can write down an integral representation for the period.

Here is how we obtain the asymptotics. For the
sequence of monic orthogonal polynomials $P_n(z)$ associated with a weight
function
$w$ on the unit circle, the square of the $L_2(w)$-norm of $P_n$ is the ratio
of two
consecutive
Toeplitz determinants associated with $w$. The asymptotics of $P_n$ gives
information
on the asymptotics of this ratio of determinants. Analogously the asymptotics
of the
continuous analogue of these polynomials, which we shall denote by $R(x)$,
determine the asymptotics of the logarithmic derivative of the Fredholm
determinant.
The underlying weight function for $R(x)$ will be the characteristic function
of the
complement of $J$ in \bR, so we have the continuous analogue of the polynomials
orthogonal on a union of arcs. The
asymptotics of orthogonal polynomials such as these were obtained in \cite{w1}
and
the ideas of this paper were used in \cite{w3} and are used here. We find an
entire function which comes close to satisfying the characteristic property of
$R$
given in (\ref{Rform}) below by first finding another function (which is not
entire,
or even single-valued) which has this property exactly (the function $h_s$ in
Lemma 4), and then we approximate this by an entire function. The entire
function we
find in this way will necessarily be a good approximation to $R$. This gives a
weaker form of (\ref{Det}) at first. It is strengthened by exploiting its
relation to
a certain extremal problem. Some details will be omitted here since the
complete
details for the analogous orthogonal polynomial case can be
found in \cite{w1}.

We obtain (\ref{Tr}) by using a representation, which is essentially contained
in
\cite{jmms}, for the resolvent kernel of $K_s$ in terms of $R(x)$.
There are analogous representations for a large
class of kernels like the sine kernel \cite{its,tw}.

In \cite{jmms} it was shown that for general $J$ the Fredholm
determinant is governed by a system of partial differential equations with the
end-points of $J$ as the independent variables. The dependent variables are
the values of the function $R(x)$ at the end-points of $J$, and
the logarithmic derivative of the Fredholm determinant has a simple
representation in
terms of these values. We do not use this representation but rather the one
alluded to above (formula (\ref{detform}) below), which makes its
asymptotics, at least by this method, more accessible. Nevertheless, our
results may
give a hint of what the asymptotics of the solutions of the system of equations
might involve.

The author wishes to thank Craig Tracy for introducing him to the subject of
random
matrices, and in particular to the problem of asymptotics for the several
interval
case.\sp

\begin{center}{\bf II. The Resolvent Kernel and the Function
\boldmath $R(x)$}\end{center}\sp

We shall denote the Fourier transform by a circumflex, as usual,
\[\hat f(x):=\int_{-\iy}^{\iy}e^{ixt}\,f(t)\,dt,\]
and write $\Es$ for the space of Fourier transforms of functions in
$L_2(-s,s)$. This consists of entire functions which are
$O(e^{s|{\rm Im}\,x|})$ and whose restriction to the real
line {\bf R} belongs to $L_2$. We also write $E$ for the complement of $J$ in
{\bf R}
and $\Om$ for the complement of $E$ in {\bf C}.
\vspace{.1in}

\noi{\bf Lemma 1}. There exists a unique function $R(x)\in e^{-isx}+\Es$ such
that
\bq\int_E\bar{(R(x)-e^{-isx})}\,g(x)\,dx=\int_Jg(x)\,e^{isx}\,dx
\quad\mbox{for all}\ g\in\Es.\label{Rform}\eq
In terms of this function we have the representations
\bq -{d\ov
ds}\log\,\det\,(I-K_s)={1\ov\pi}|J|+{1\ov\pi}\int_E|R(x)-e^{-isx}|^2\,dx,
\label{detform}\eq
where $|J|$ is the measure of $J$, and
\bq\mbox{tr}\;K_s\,(I-K_s)^{-1}=-{1\ov
\pi}\,\im\int_JR'(x)\,\bar{R(x)}\,dx.\label
{trform}\eq \sp

\noi{\bf Remark}. The set $e^{-isx}+\Es$ is analogous to the set of monic
polynomials of a given degree, and (\ref{Rform}) implies that $R(x)$
is orthogonal on $E$ to the Fourier transforms of a dense set of
functions in $L_2(-s,\,s)$, those which are smooth and vanish at $-s$. (The
integral
expressing the orthogonality is a principal value integral at infinity.) Thus
$R(x)$ is analogous to
a monic polynomial of a family orthogonal on a union of circular arcs. The
identity
(\ref{detform}) is the analogue of the fact that the square of the norm of the
monic
polynomial is the ratio of two consecutive determinants.\sp

\noi{\bf Proof}. Write
\[K_s(x,y):={\sin s\,(x-y)\ov\pi(x-y)},\]
and  denote by $K_s$ the operator on all of {\bf R} with kernel
$K_s(x,y)\,\ch_J(y)$.
Our determinant and trace are the same for this kernel as for the other. We
express the resolvent kernel $R_s(x,y)$ for this operator in terms of the
functions
\[R_{\pm}(x):=(I-K_s)^{-1}\,e^{\pm isx}\]
introduced in \cite{jmms}. If $M$ denotes multiplication by $x$
then the commutator $[K_s,\,M]$ has kernel
\[{1\ov 2\pi i}(e^{isx}\,e^{-isy}-e^{-isx}\,e^{isy}).\]
It follows from this upon left- and right-multiplying by $(I-K_s)^{-1}$ that
for
$x,y\in J$
\[R_s(x,y)={R_+(x)\,R_-(y)-R_-(x)\,R_+(y)\ov2\pi i(x-y)},\]
and in particular
\[R_s(x,x)={R_+'(x)\,R_-(x)-R_-'(x)\,R_+(x)\ov2\pi i}.\]

We write $R(x)$ for $R_-(x)$ and observe that, since the kernel of $K_s$ is
real,
$R_+(x)=\bar{R(x)}$. Thus
\[R_s(x,y)={\bar{R(x)}\,R(y)-R(x)\,\bar{R(y)}\ov2\pi i(x-y)},\]
\[R_s(x,x)=-{1\ov \pi}\,\im R'(x)\,\bar{R(x)}.\]
This establishes (\ref{trform}).
Next, since the kernel of $K_s'$ (differentiation is with respect to $s$) is
\[{1\ov2\pi}(e^{is(x-y)}+e^{-is(x-y)})\ch_J(y),\] the kernel of
$(I-K_s)^{-1}\,K_s'$ equals
\[{1\ov2\pi}(R_+(x)\,e^{-isy}+R_-(x)\,e^{isy})\,\ch_J(y)={1\ov\pi}\,\re
R(x)\,e^{isy}\,
\ch_J(y).\]
Thus, since
\[{d\ov ds}\log\,\det\,(I-K_s)=-\mbox{tr}\;(I-K_s)^{-1}\,K_s',\]
we obtain
\bq-{d\ov ds}\log\,\det\,(I-K_s)={1\ov \pi}\,\re\int_JR(x)\,e^{isx}\,dx.
\label{drep}\eq
If we apply the identity (\ref{Rform}), which will be proved momentarily, to
the
function $g(x)=R(x)-e^{-isx}$ it gives
\[\int_E|R(x)-e^{-isx}|^2\,dx=\int_JR(x)\,e^{isx}\,dx-|J|,\]
and (\ref{detform}) follows from (\ref{drep}).

To obtain the first assertion of the lemma observe first that the difference
between
any two functions $R$ satisfying (\ref{Rform}) belongs to $\Es$ and is
orthogonal
on $E$ to $\Es$, and so it vanishes on $E$. Since $\Es$ consists of
entire functions, this
difference must be 0. To show that our fuction $R(x)=R_-(x)$ satisfies
(\ref{Rform}) we write its definition in the equivalent form
\[\bar{R(x)}-\int_JK_s(x,y)\,\bar{R(y)}\,dy=e^{isx}\]
and observe that $K_s(x,y)$ is the kernel of the projection operator from
$L_2({\bR})$
to $\Es$. Hence $g\in\Es$ implies
\[\int_{\bR}g(x)\,K_s(x,y)\,dx=g(y).\]
Thus if also $g\in L_1(\bR)$ then multiplying both sides of the previous
identity by
$g(x)$ and integrating over {\bf R} give
(\ref{Rform}) in this case. The extension to general $g$ is
straightforward.\hspace{.2in}$\Box$\sp

\begin{center}{\bf II. Green Functions, Neumann functions and\\
a Reproducing Function for \boldmath $E$}\end{center}\sp

In \cite{w3}, where $J=[-1,1],\,E=(\iy,-1]\cup[1,\iy)$ and $\Om=\bC\backslash
E$, the
procedure was to show first that (\ref{Rform}) holds if $R(x)$ is replaced by
the
sum of the two limiting values on $E$ of the function
\bq
h(x):=e^{-is\sqrt{x^2-1}}\sqrt{{x+\sqrt{x^2-1}\ov2\sqrt{x^2-1}}}\label{m=1}\eq
defined in $\Om$ with appropriate branches of the square roots. Notice that
$R(x)$ and
$h(x)$ have the same asymptotic behavior as $x\ra\iy$ in $\Om$. One
approximates
$h(x)$ by an entire function, for which (\ref{Rform}) holds approximately, and
then
deduces that this function must be a good approximation to $R(x)$.

The function $\sqrt{x^2-1}$ in (\ref{m=1}) was chosen to be that branch of the
square root which is analytic in $\Om$ and asmptotically equal to $x$ as
$x\ra\iy$ in the upper half-plane. Its imaginary part vanishes on $E$ and
equals
$|\im x|+O(1)$ in $\Om$, and so is a kind of Green function for $\Om$. We begin
by
constructing its analogue for general
\[J=\bigcup_{k=1}^m\,[\al_k,\,\be_k].\]
Let
\[q(x)=\prod_ {k=1}^m\,(x-\al_k)\,(x-\be_k)\]
and let $\sqrt{q(x)}$ denote that branch which is analytic in $\Om$ and
asymptotic to
$x^m$ as $x\ra\iy$ in the upper half of $\Om$. This has purely real limiting
values
on $E$. Let $p(x)$ be the monic $m$'th degree polynomial determined by the $m$
equations
\bq\int_{\al_k}^{\be_k}{p(y)\ov \sqrt{q(y)}}\,dy=0,\qquad(k=1,\cdots,m).\label
{conds}\eq
Then define
\bq G_0(x):=\int_{\al_1}^x{p(y)\ov\sqrt{q(y)}}\,dy,\qquad(x\in\Om).\label{G}\eq
This is multiple-valued when $m>1$ because of the presence of the intervals
$[\be_k,\al_{k+1}]$\linebreak $(k\leq m-1)$ one might integrate around and get
a
nonzero result.
If $C_k$ is a curve going around this interval and none of the others then the
periods
\bq\De_{C_k}\,G_0=\oint_{[\be_k,\,\al_{k+1}]}{p(y)\ov\sqrt{q(y)}}\,dy\label{periods}\eq
are purely real, so that $\im G(x)$ is a single-valued harmonic function
in $\Om$. The conditions (\ref{conds}) guarantee that $\im G(x)\ra 0$ as $x\ra
E$.

For some real constants $a_1'$ and $a_2$ we have
\[{p(x)\ov\sqrt{q(x)}}=1+{a_1'\ov x}+{a_2\ov x^2}+O({1\ov x^3})\quad
{\rm as}\;x\ua\iy.\]
Here the notation $x\ua\iy$ means that $x\ra\iy$ in the upper half-plane of
$\Om$.
(Similarly $x\da\iy$ will mean that $x\ra\iy$ in the lower half-plane of
$\Om$.)
The integral
\[\int_{-\iy+0i}^{\iy+0i}\left({p(x)\ov\sqrt{q(x)}}-1\right)\,dx\]
equals $-\pi\,i\,a_i'$. But it follows from (\ref{conds}) that it is also real.
Hence
$a_1'=0$ and so
\bq{p(x)\ov\sqrt{q(x)}}=1+{a_2\ov x^2}+O({1\ov x^3})\quad{\rm as}\;x\ua\iy.
\label{pq}\eq
It follows readily that
\[G_0(x)=x+\left[\int_{\al_1}^{\iy+0i}\left({p(y)\ov\sqrt{q(y)}}-1\right)\,dy-\al_1\right]
-{a_2\ov x}+O({1\ov x^2}),\]
where $a_1$, the expression in brackets, is purely real. We now define
\[G(x):=G_0(x)-a_1.\]
$G(x)$ has the same Green function
characteristics as $G_0(x)$ and has the behavior
\bq G(x)=x-{a_2\ov x}+O({1\ov x^2})\quad{\rm as}\;x\ua\iy.\label{Gasymp1}\eq
Since $G_0$ is purely imaginary on $J$, we have $G_0(\bar{x})=-\bar{G_0(x)}$,
and so from the above we deduce
\bq G(x)= -x-2\,a_1+{a_2\ov x}+O({1\ov x^2})\quad{\rm
as}\;x\da\iy.\label{Gasymp2}\eq

The function $e^{isG(x)}$ is multiple-valued and analytic in $\Om$ and has
single-valued
absolute value. Such a function can be thought of as a section of a holomorphic
line
bundle over $\Om$, but instead we proceed as follows. Denote
by $\cH$ the set of all (single-valued) analytic functions $f$ in the
complement of
$(-\iy,\al_1]\cup[\be_1,\iy)$ in $\bC$ for which the limits
\bq \lim_{\ve\ra0+}{f(x+i\ve)\ov f(x-i\ve)},\qquad x\in[\al_k,\be_k],
\quad (k=2,\cdots,m) \label{lims}\eq
are constants of absolute value 1. Each such function continues to a
multiple-valued
function in $\Om$ with single-valued absolute value. Two functions in $\cH$ are
said to
belong to the same class if the corresponding limits (\ref{lims}) for them are
the
same. A class is denoted by $\Ga$ and the set of classes can be identified with
the
$m-1$-torus. We denote by $\cH(\Ga)$ the functions of class $\Ga$. Occasionally
we
shall allow our functions to have poles. An example is the exponential Green
function
$\Ph(x,x_0)$ described below. We shall write our classes additively, and so
each will
be an element of ${\bf T}^{m-1}:=(\bR/2\pi{\bf Z})^{m-1}$. If $\Ga_0$ is the
class
of $e^{iG(x)}$ then $s\Ga_0$ is the class of $e^{isG(x)}$. This is our curve in
${\bf T}^{m-1}$.

Next we have to find an analogue of the second factor in (\ref{m=1}). There
will be
many, one for each class $s\Ga_0$, and so we will find one for a general class
$\Ga$.
We begin by recalling the characteristic properties of the Green and Neumann
functions
for $\Om$. Given $x_0$, Green's function with pole at $x_0$, denoted by
$g(x,x_0)$,
is harmonic for $x\in\Om$, except at $x_0$ where it has the singularity
$\log|x-x_0|
^{-1}$, and has limit 0 as $x\ra E$. Denoting by $\tilde{g}(x,x_0)$ its
(multiple-valued)
harmonic conjugate, we define the exponential Green function by
\[\Ph(x,x_0):=e^{g(x,x_0)+i\tilde{g}(x,x_0)}.\]
Neumann's function $N(x,x_0,x_1)$ is harmonic except at $x=x_0$ and $x=x_1$,
where
it has the singularities $\log|(x-x_1)/(x-x_0)|$, and on $E$
\[{\pl N(x,x_0,x_1)\ov\pl n_x}=0.\]
Here $\pl/\pl n_x$ denotes the two derivatives normal to $E$. We set
\[\Ps(x,x_0,x_1):=e^{N(x,\,x_0,\,x_1)+i\tilde{N}(x,\,x_0,\,x_1)}.\]
This has a simple pole at $x=x_0$ and a simple zero at $x=x_1$, unless the two
are
equal, in which case the function equals 1.
The function $\Ph(x,x_0)$ is multiple-valued with single-valued absolute value,
and
is determined only up to an arbitrary constant factor of absolute value 1. The
function $\Ps(x,x_0,x_1)$ is single-valued but determined only up to an
arbitrary
nonzero constant factor. The analogue of the second factor in (\ref{m=1}) will
be
built out of these functions.\sp

\noi{\bf Lemma 2}. The function $1+p(x)/q(x)^{\hf}$ is nonzero (i.e., has
nonzero
limits) on $E$ and has $m-1$ zeros in $\Om$.\sp

\noi{\bf Proof}. We know that
\bq \im(x+G(x))=0\ {\rm on}\ E.\label{im1}\eq
{}From (\ref{Gasymp1}) and (\ref{Gasymp2}) we deduce that
\[\underline{\lim}\,_{x\ra\iy}\im(x+G(x)\geq0.\]
Therefore
\bq \im(x+G(x))>0\ {\rm in}\ \Om.\label{im2}\eq
If
\[1+{p(x)\ov\sqrt{q(x)}}=(x+G(x))'\]
had limit zero at some point of $E$ this would contradict (\ref{im1}) and
(\ref{im2})
since at a critical point an analytic function maps a local half-disc to a full
disc.

For the second statement let $C_k$ be the contours in (\ref{periods}) and $C
=C_-\cup C_+$, where $C_-$ starts at $-\iy-0i$, goes around $\al_1$, and ends
at
$-\iy+0i$, while $C_+$ starts at $\iy+0i$, goes around $\be_m$, and ends at
$\iy-0i$. All the contours are described so that the parts below $E$ are
traversed to
the right, the parts above $E$ to the left. We find that
\[\De_C\arg\left(1+{p(x)\ov\sqrt{q(x)}}\right)=0,\quad\De_{C_k}\arg\left(1+{p(x)
\ov\sqrt{q(x)}}\right)=-2\pi,\;(k=1,\cdots,m-1).\]
This and the argument principle establish the lemma.\hspace{.2in}$\Box$\sp

Denote the zeros of $1+p(x)/q(x)^{\hf}$ by $x_1\st,\cdots,x_{m-1}\st$. It
follows from
a discussion in \cite{w1} that for each class $\Ga$ there are unique points
$x_1,
\cdots,x_{m-1}$, and unique numbers $\ve_1,\cdots,\ve_{m-1}$ each equal to
$\pm1$,
such that
\bq\prod_{k=0}^{m-1}\Ps(x,x_k\st,x_k)\geq0\ {\rm on}\ E\label{pos}\eq
(with an appropriate normalization of the $\Ps$'s) and such that
\bq k_{\Ga}(x):=\left[\hf\left(1+{p(x)\ov\sqrt{q(x)}}\right)\prod_{k=0}^{m-1}
\Ps(x,x_k\st,x_k)\,\Ph(x,x_k)^{\ve_k}\right]^{\hf}\in\Ga.\label{k}\eq
The $x_k$ are given by the solution of a certain Jacobi inversion problem. The
details,
which we shall not present, can be found in \S6 of \cite{w1}.

One possibility, which occurs for certain $\Ga$, is that all $x_k=x_k\st$ so
the
$\Ps$'s don't appear at all. In general, though, they do appear and the zeros
of
$1+p(x)/q(x)^{\hf}$ are cancelled by the poles of the product of the $\Ps$'s.
The
zero $x_k$ of $\Ps(x,x_k\st,x_k)$ is cancelled by the pole of
$\Ph(x,x_k)^{\ve_k}$ if
$\ve_k=1$ and reinforced to a double zero if $\ve_k=-1$. So there is no extra
multiple-valuedness introduced by taking the square root. Observe, though, that
we have
not quite defined $k_{\Ga}$ because of the nonuniqueness of the exponential
Green and
Neumann functions. We make them unambiguous by the requirements
\bq N(x,x_k\st,x_k),\, \tl{N}(x,x_k\st,x_k),\, \tl{g}(x,x_k\st,x_k)\,\ra 0\quad
\mbox{as}\ x\ua\iy,\label{normalizations}\eq
and then choose the square root so that $k_{\Ga}(x)\ra1$ as $x\ua\iy$.

Our eventual replacenent for $h(x)$ will be
\bq h_s(x):=e^{-isG(x)}\,k_{s\Ga_0}(x).\label{hs}\eq\sp

\noi{\bf Lemma 3}. Suppose $\ps\in\cH(\Ga)$ has the behavior
\[\ps(x)={c\ov x}+O({1\ov x^2})\ {\rm as}\ x\ua\iy,\qquad
\ps(x)= O({1\ov x})\ {\rm as}\ x\da\iy.\]
Then
\[\oint_E\overline{k_{\Ga}(x)}\,\ps(x)\,|dx|=-\pi\,i\,c.\]\sp

\noi{\bf Remark 1}. The conjugate-analytic function $\overline{k_{\Ga}}$
belongs to
the class $-\Ga$ so that $\overline{k_{\Ga}(x)}\,\ps(x)$ is single-valued in
$\Om$.
The notation $\oint_E\cdots|dx|$ indicates that $E$ is not oriented but is
taken
twice, using the two limiting values of the function.\sp

\noi{\bf Remark 2}. It is because of this caracteristic property of $k_{\Ga}$,
reminiscent of that of reproducing kernels, that we call it the ``reproducing
function''
associated with $\Ga$ and $\Om$: integrating $\ps$ against $\overline{k_{\Ga}}$
yields the constant determining $\ps$'s behavior as $x\ua\iy$.\sp

\noi{\bf Proof}. Let $j_{\Ga}(x)$ be the function obtained by replacing each
$\ve_k$
by $-\ve_k$ in (\ref{k}). Then
\bq |j_{\Ga}(x)|=|k_{\Ga}(x)|\ {\rm on}\ E\label{jk1}\eq
and, by (\ref{pos}),
\bq j_{\Ga}(x)\,k_{\Ga}(x)\ \mbox{is real-valued on}\ E\ \mbox{with the same
sign as}\
1+{p(x)\ov\sqrt{q(x)}}.\label{jk2}\eq
{}From the fact that $\im(x+G(x))$ is zero on $E$ and positive in $\Om$ it
follows that
\[1+{p(x)\ov\sqrt{q(x)}}=(x+G(x))'\]
has positive limit on $E$ from above and negative limit from below.
Alternatively, if
we denote the ``upper'' (resp. ``lower'') part of $E$ by $E^+$ (resp. $E^-$),
then
\[{\rm sgn}\,\left(1+{p(x)\ov\sqrt{q(x)}}\right)=\left\{
\begin{array}{ll}1&{\rm on}\ E^+\\-1&{\rm on}\ E^-.\end{array}\right.\]
Hence, from (\ref{jk1}) and (\ref{jk2})
\[k_{\Ga}\,\overline{k_{\Ga}}=k_{\Ga}\,j_{\Ga}\times\left\{
\begin{array}{ll}1&{\rm on}\ E^+\\-1&{\rm on}\ E^-,\end{array}\right.\]
since both sides have the same absolute value and are positive on $E$. Dividing
by
$k_{\Ga}$ we see that the integral in the statement of the lemma equals
\[\oint_Ej_{\Ga}(x)\,\ps(x)\,dx,\]
where now $E$ is oriented so that $E^+$ is traversed to the right, $E^-$ to the
left.
Observe that $j_{\Ga}\in\cH(-\Ga)$ and so $j_{\Ga}(x)\,\ps(x)$ extends to a
single-
valued analytic function in $\Om$. Moreover
\bq j_{\Ga}(x)\,\ps(x)=\left\{\begin{array}{ll}O(x^{-2})&{\rm as}\
x\da\iy,\\&\\
c\,x^{-1}+O(x^{-2})&{\rm as}\ x\ua\iy\end{array}\right.\label{jps}\eq
(the first because $1+p(x)/\sqrt{q(x)}=O(1/x)$ as $x\da\iy$). The integral is
equal to
the limit of the sum of the integrals over two
semicircles, one in the upper half-plane and one in the lower, as their radii
tend to
infinity. By (\ref{jps}) the integral over the upper semi-circle tends to 0
while the
integral over the lower tends to $-\pi\,i\,c$.\hspace{.3in}$\Box$\sp

Now recall that we define $h_s(x)$ by (\ref{hs}). \sp

\noi{\bf Lemma 4}. Suppose $g(x)$ is single-valued and analytic in $\Om$ and
satisfies
\[e^{isx}g(x)={c\ov x}+O({1\ov x^2})\ \mbox{as}\ x\ua\iy,\qquad
e^{-isx}g(x)=O({1\ov x})\ \mbox{as}\ x\da\iy.\]
Then
\[\oint_E\bar{h_s(x)}\,g(x)\,|dx|=-\pi\,i\,c.\]\sp

\noi{\bf Proof}. Write the intgral as
\[\oint_E\bar{k_{s\Ga_0(x)}}\,e^{isG(x)}\,g(x)\,|dx|.\]
The second factor belongs to class $s\Ga_0$ and by (\ref{Gasymp1}) and
(\ref{Gasymp2})
has the properties of $\ps(x)$ in the statement of Lemma 3. The assertion
follows.
\hspace{.2in}$\Box$\sp

\begin{center}{\bf III. Asymptotics of \boldmath$R(x)$} \end{center}\sp

Define, for $x\in E$,
\[h_{\pm}(x):=\lim_{\ve\ra0\pm}h_s(x+i\ve).\]
We shall deduce from Lemma 4 and the first statement of Lemma 1 that $R$ is
well-approximated on $E$ by $h_++h_-$. To do this we first replace
$h_+(x)+h_-(x)$ by an entire function with the same general behavior as $R$ and
which is close to $h_++h_-$ on $E$. Such a function will be $e^{-isx}+q(x)$,
where
\bq q(x):={e^{-isx}\ov 2\pi i}\int_{ia-\iy}^{ia+\iy}{e^{isy}\,h(y)\ov y-x}\,dy
\label{q}\eq
with $a<{\rm min}\;(0,\,\im x)$.
(We have dropped the subscript $s$ in the notation for the function $h_s$.) To
prove
that we do get a good approximation we derive an integral
representation for $q(x)$ which requires deforming the path of integration in a
rather
involved way, as well as the analytic continuation of $h(y)$.

Recall that the exponential Green functions in (\ref{k}) have absolute value 1
on $E$
and by (\ref{pos}) the the product of exponential Neumann functions is positive
there.
Thus both of these continue to the Riemann surface obtained by joining $\Om$ to
a copy
$\bar{\Om}$ in the usual way, by identifying
$E^{\pm}$ in $\Om$ with $E^{\mp}$ in $\bar{\Om}$.
Similarly $G(y)$ continues into $\bar{\Om}$ and its
continuation has {\em negative} imaginary part there. The resulting
continuation of $h(y)$ is not single-valued because of the presence of branch
points
at the $\al_k$ and $\be_k$. We shall deform our path of integration to a
system of contours, all lying in $\bar{\Om}$ where $\im G(y)>0$, and all lying
in the
upper half-plane where the factor $e^{isy}$ is small.

Before stating the main lemma we describe the contours $A_i,\;(i=1,\cdots,2m)$.
If
$i=2k-1$, then $A_i$ begins at $\al_k+i\iy$ and goes down toward $\al_k$; it
makes
1\,1/2 infinitesimal counterclockwise circuits of $\al_k$ and then goes back up
to
$\al_k+i\iy$. If $i=2k$ then $A_i$ begins and ends at $\be_k+i\iy$ but goes
around $\be_k$, when it gets to it, clockwise instead of counterclockwise.
Except for
the infinitesimal loops around the points of $\pl J$, these contours lie
entirely
in the upper half-plane of $\bar{\Om}$, and so have the sought properties.

The function $h(y)$ in the integrand below will be a continuation of the
original $h(y)$.
(This will be clear during the course of the derivation.) It can be described
as that
continuation of $h$ which, at the initial points of the contours $A_i$, is the
function obtained by continuing our original $h$ from below $E$ across it into
the
upper half-plane; on the rest of the $A_i$ we take the continuation along $A_i$
of
this one. Observe that when our contours first cross the real line through $E$,
as
they do, $h(y)$ returns to its original value since we will have undone the
first
continuation. \sp

\noi{\bf Lemma 5}. For $x\in E$ we have
\bq e^{-isx}+q(x)=h_+(x)+h_-(x)+e^{-isx}\sum_{i=1}^{2m}{(-1)^{i+1}\ov2\pi i}
\int_{A_i}{e^{isy}\,h(y)\ov y-x}\,dy.\label{qrep}\eq\sp

\noi{\bf Proof}. The path of integration in (\ref{q}) may be deformed to the
union of
the contours $C$ and $C_k$ which occur in the proof of Lemma 2 (all described
so that
$E^-$ is traversed to the right and $E^+$ to the left) plus an infinitely large
semi-circle in the upper half-plane traversed clockwise. Since $e^{isy}h(y)=
1+O(|y|^{-1})$ as $y\ua\iy$ the integral over the semi-circle equals $-\pi i$,
and so
\bq e^{isx}\,q(x)=-\hf+{1\ov2\pi i}\int_{C\cup C_1\cup\cdots
C_{m-1}}{e^{isy}\,h(y)
\ov y-x}\,dy.\label{q1}\eq
Suppose now, for definiteness, that $x\in(-\iy,\,\al_1)$, and consider $C_-$,
the left
part of $C$, which goes from $-\iy-0i$, around $\al_1$ counterclockwise, and
then to
$-\iy+0i$. We move the upper part of $C_-$ through $(-\iy,\,\al_1)$ into the
lower
half-plane and the lower part of $C_-$ through $(-\iy,\,\al_1)$ into the upper
half-plane. The result is a contour which runs from $\al_1+i\iy$ to $\al_1$,
around
$\al_1$ counterclockwise, and then
down to $\al_1-i\iy$, with $h(y)$ in the integrand replaced by its
continuation. If we
keep in mind how $G(y)$  and $k_{\Ga}$ continue we see that the continuation of
$h(y)$
from the lower half-plane to the upper satisfies
\[e^{isy}\,h(y)=O({1\ov|y|}),\]
while its continuation from the upper to the lower satisfies
\[e^{isy}\,h(y)=1+O({1\ov|y|}).\]
This shows that the deformation is valid if we add to the integral the integral
over
an infinitely large quarter-circle in the lower half-plane, which is $-\pi
i/2$,
as well as the contributions
from the pole at $y=x$, which is passed twice during the deformation. It
follows from
this discussion that if we replace $C_-$ by this new contour, which we call
$B_1$,
then we must add
\[-{1\ov4}+e^{isx}\,h_+(x)+e^{isx}\,h_-(x)\]
to the right side of (\ref{q1}). Next, the contour $C_1$, which runs
counterclockwise
around $[\be_1,\,\al_2]$, we deform by pushing its upper part down through
$(\be_1,\,\al_2)$
into the lower half-plane and its lower part up through $(\be_1,\,\al_2)$ into
the
upper half-plane. Both continuations of $e^{isy}\,h(y)$ are bounded at $\iy$
and the
result is that we can replace $C_1$ by $-B_2+B_3$, where $B_2$ (resp. $B_3$)
starts at
$\be_1+i\iy$ (resp. $\al_2+i\iy$), loops around $\be_1$ clockwise (resp.
$\al_2$
counterclockwise), and ends at $\be_1-i\iy$ (resp. $\al_2-i\iy$).
During this deformation no extra terms are picked up. We continue analogously
with
the remaining contours $C_k$ and then end with $C_+$, the right part of $C$,
deforming
it as we did $C_-$. [The reader is advised to draw a
picture.] Another integral over an infinitely large quarter-circle is picked up
during this last deformation. The result at this stage is
\bq e^{isx}\,q(x)=-1+e^{isx}\,h_+(x)+e^{isx}\,h_-(x)+\sum_{i=1}^{2m}
{(-1)^{i+1}\ov2\pi i}\int_{B_i}{e^{isy}\,h(y)\ov y-x}\,dy.\label{q2}\eq
Finally the lower parts of $B_1$ and
$-B_2$ can be joined into a single contour, and the part of this new contour
which runs from $\al_1$ to $\be_1$ in the lower half-plane can be pushed up
through
$(\al_1,\,\be_1)$ into the
upper half-plane, resulting in the contour $A_1- A_2$ as in the statement of
the
lemma. This deformation is valid since $G(x)$ continues to have negative
imaginary part
during it. The remaining contours are deformed similarly, and this completes
the
demonstration of the lemma.\hspace{.2in}$\Box$\sp

\noi{\bf Lemma 6}. We have, as $s\ra\iy$,
\[\int_E|e^{-isx}+q(x)-h_+(x)-h_-(x)|^2\,dx=O(s^{-1}).\]\sp

\noi{\bf Proof}. We must show that the norm in $L_2(E)$ of the sum in
(\ref{qrep}) is $O(s^{-\hf})$. Consider the integral over $A_1$, which can be
written
as a single integral paramaterized as $y=\al_1+i\,t,\;t\geq0$. (The integral
over the
infinitesimal loops around $\al_1$ vanish.) For each $\dl>0$ the
integral over $t\geq\dl$ is bounded by a quantity which is exponentially small
in $s$,
uniformly
in $x$, times $O((1+|x|)^{-1})$, so the $L_2(E)$ norm of this part of the
integral
is exponentially small. In the neighborhood of $t=0$, $\im G(\al_1+it)$ is
asymptotically
a {\em negative} constant times $t^{\hf}$ (see the remark below), while
\bq |k_{s\Ga_0}(\al_1+it|=O(t^{-1/4}).\label{hest}\eq
(This last comes from the behavior of $1+p/\sqrt{q}$.) Since on $A_1$
\[{1\ov|x-y|}=O\left({1\ov |x-\al_1|+t}\right),\]
we see that this part of the integral is bounded by a constant times
\bq\int_0^{\dl}{e^{-\eta s\sqrt{t}}\ov
t^{1/4}\,(|x-\al_1|+t)}\,dt\label{est}\eq
for some $\eta>0$ and it is a simple exercise to show that the $L_2({\bf R})$
norm of
this is $O(s^{-\hf})$. The other $A_i$ are treated in a similar manner.
\hspace{.2in}$\Box$\sp

\noi{\bf Remark}. Since the function $k_{s\Ga_0}$ in (\ref{hest}) depends on
the
parameter $s$ we have to
know that the estimate holds uniformly in $s$. The reason it does is that the
Green
and Neumann functions appearing in (\ref{k}) are smooth, uniformly in their
parameters,
away from their poles in the Riemann surface $\Om\cup\bar\Om$. (See, for
example,
Lemma 4.1 of \cite{w1}.) The local coordinate for the surface near $x=\al_1$ is
$(x-\al_1)^{\hf}$ and it follows that $k_{s\Ga_0}$ and $G(x)$ are smooth
functions of
$(x-\al_1)^{\hf}$. This shows that (\ref{hest}) holds
uniformly in $s$ and also accounts for the $\sqrt{t}$ in the exponent of the
integrand
in (\ref{est}).\sp

\noi{\bf Lemma 7}. We have, as $s\ra\iy$,
\[\int_E|R(x)-h_+(x)-h_-(x)|^2\,dx=O(s^{-1}).\]\sp

\noi{\bf Proof}. The integrand equals
\[(R(x)-h_+(x)-h_-(x))\,\bar{(e^{-isx}+q(x)-h_+(x)-h_-(x))}\]
\[+(R(x)-h_+(x)-h_-(x))\,\bar{(R(x)-e^{-isx}-q(x))}.\]
Let us look at the last term. It follows from Lemmas 1 and 4 that the integral
over $E$
of the first factor times the complex conjugate of the Fourier transform of any
smooth
function in $L_2(E)$ vanishes, and by continuity that Fourier transform does
not have
to be of a smooth function. Now $R(x)-e^{-isx}$
is the Fourier transform of a function in  $L_2(E)$, and so is $q(x)$: it is
easily
seen from its definition (\ref{q}) that $q(x)$ is an entire function of
exponential
type $s$
and it follows from Lemma 6 that it belongs to $L_2({\bf R})$. Thus the
integral over
$E$ of the second term above vanishes. By Lemma 6 and Schwarz's inequality the
integral
of the first term is $O(s^{-\hf})$ times the square root of the integral we
started
with. This establishes the lemma.\hspace{.2in}$\Box$\sp

This is all we shall need for (\ref{Det}) and the reader interested
only in these can go on to the next section. A few more estimates will be
needed for the
proof of (\ref{Tr}).\sp

\noi{\bf Lemma 8}. (a) $e^{-isx}+q(x)=h(x)+O(s^{-3/2})$ uniformly on compact
subsets of $\Om$.\\
(b) For every $\ve>0$ there exists a neighborhood of $\pl J$ in which $q(x)=
O(e^{\ve s})$.\sp

\noi{\bf Proof}. There is a modification of (\ref{qrep}), proved in the same
way,
which holds for $x\in\Om$: the
sum $h_+(x)+h_-(x)$ is replaced by $h(x)$ and the contours $A_i$ are adjusted
so as to avoid $x$ (or the compact subset of $\Om$ in which they may lie), but
they
still lie in the upper half-plane of $\bar{\Om}$. This is easily seen to give
assertion
(a). It follows as a special case of this that if $\dl$ is sufficiently  small
then
$q(x)=O(e^{\ve s})$ on the circular arc
\[x=\al_k+\dl\,e^{i\theta},\quad -{3\pi\ov4}<\theta<{3\pi\ov 4}.\]
For the rest of the circle we use the continuation of (\ref{qrep}),
as originally stated, from $x\in E$ onto that arc and obtain a similar
estimate. Since
$q$ is analytic inside the circle it is $O(e^{\ve s})$ in the full disc. A
similar
argument applies, of course, to neighborhoods of the $\be_k$.
\hspace{.2in}$\Box$\sp

\noi{\bf Lemma 9}. (a) $R(x)=h(x)+O(s^{-\hf}\,e^{s\,\im G(x)})$ uniformly on
compact
subsets of the interior of $J$.\\
(b) For every $\ve>0$ there exists a neighborhood of $\pl J$ in ${\bf C}$ in
which
$R(x)=O(e^{\ve s})$.\sp

\noi{\bf Proof}. The function
\bq\eta(x):=|(R(x)-e^{-isx}-q(x))\,e^{isG(x)}|^2\label{eta}\eq
is bounded and subharmonic in $\Om$ and it follows from Lemmas 6 and 7 that its
boundary
function satisfies
\[\oint \eta(x)\,|dx|=O(s^{-1}).\]
Hence, from general considerations, $\eta(x)=O(s^{-1})$ uniformly on compact
subsets of $\Om$. Combining this with the estimate of Lemma 8(a) proves our
first
assertion. (Observe that the error term in the statement is exponentially large
and so
dominates the term $e^{-isx}$.) To prove the second we replace $G(x)$ in
(\ref{eta}) by the analogous
function where $J$ is replace by $J$ minus small neighborhoods of its boundary
points. We then get an estimate for $R(x)-e^{-isx}-q(x)$ also in a neighborhood
of
$\pl J$, apply Lemma 9(b), and deduce the second assertion.
\hspace{.2in}$\Box$\sp

\begin{center}{\bf IV. Demonstration of (\ref{Det})} \end{center}\sp

Let us first evaluate asymptotically the integral
\bq\cI:=\int_E|h_+(x)+h_-(x)-e^{-isx}|^2\,dx.\label{hint}\eq
It can be written as
\[\int_E\bar{h_+(x)}\,(h_+(x)-e^{-isx})\,dx+\int_E\bar{h_-(x)}\,(h_-(x)-e^{-isx})\,dx\]
\[-\int_E(h_+(x)\,e^{isx}+h_-(x)\,e^{isx}-1)\,dx+\int_E(\bar{h_+(x)}\,h_-(x)
+\bar{h_-(x)}\,h_+(x))\,dx.\]
Since this is real it equals the real part of what is obtained by replacing the
third
integral by its complex conjugate, so the above equals
\[\re\left\{\oint_E\bar{h(x)}\,(h(x)-e^{-isx})\,dx
-\int_E(\bar{h_+(x)}\,e^{-isx}+\bar{h_-(x)}\,e^{-isx}-1)\,dx\right\}\]
\[+2\,\re\int_E\bar{h_+(x)}\,h_-(x)\,dx=\re(\cI_1+\cI_2),\]
say. Thus $\cI=\re(\cI_1+\cI_2).$ We shall evaluate $\cI_1$ exactly and show
that
$\cI_2$ is $O(s^{-1})$.\sp

\noi{\em Evaluation of} $\cI_1$:\sp

We write our integrals over $E$ as limits as $r\ra\iy$ of the corresponding
integrals
over $E\cap[-r,\,r]$, which can be combined. We obtain
\bq\cI_1=\lim_{r\ra\iy}\left\{\oint_E\bar{h(x)}\,(h(x)-2\,e^{-isx})\,dx+2\,r\right\}
-|J|.\label{cI1}\eq
Now we can almost, but not quite, apply Lemma 4 here with
\[g(x)=h(x)-2\,e^{-isx}.\]
The problem is that this function does not satisfy the first hypothesis of the
lemma,
but we shall just modify its proof. Recall the notation there;
for convenience we drop all subscripts $\Ga$. We have
\[\oint_E\bar{h(x)}\,(h(x)-2\,e^{-isx})\,dx
=\oint_Ej(x)\,e^{isG(x)}\,(e^{-isG(x)}\,k(x)-2\,e^{-isx})\,dx\]
\[=\int_{C_r^+\cup C_r^-}j(x)\,(k(x)-2\,e^{is(G(x)-x)})\,dx,\]
where $C_r^+$ and $C_r^-$ are semi-circles described clockwise in the upper and
lower
half-planes, respectively, joining $-r$ and $r$. Since $j(x)$ and $k(x)$ are
$O(x^{-1})$ as $x\da\iy$ and $\im (G(x)-x)>0$ there, the integral over $C_r^-$
tends
to zero as $r\ra\iy$. As for $C_r^+$, both $j(x)$ and $k(x)$ tend to 1 as
$x\ua\iy$,
and are even analytic there. Suppose
\[j(x)=1+{j_0\ov x}+O({1\ov x^2}),\quad k(x)=1+{k_0\ov x}+O({1\ov x^2})\qquad
\mbox{as}\ x\ua\iy.\]
If we use (\ref{Gasymp1}) then we find that the integrand equals
\[-1+{2ia_2s\ov x}+{k_0-j_0\ov x}+O({1\ov x^2})\]
as $x\ua\iy$. Hence the integral over $C_r^+$ equals
$-2r+\pi\,(2a_2s+i(j_0-k_0))
+o(1)$ as $r\ra\iy$ and (\ref{cI1}) gives
\[\cI_1=2 \pi a_2 s+\pi i(j_0-k_0)-|J|,\]
whence
\[\re\cI_1=2 \pi a_2 s+\pi\,\im(k_0-j_0)-|J|.\]
Observe that
\[k_0=\lim_{x\ua\iy}x\,(k(x)-1)=\lim_{x\ua\iy}x\,\log k(x),\]
and similarly for $j(x)$. Thus
\[\im(k_0-j_0)=\lim_{x\ra\iy+0i}x\,\arg{k(x)\ov j(x)}.\]
Now $j(x)$ is obtained from $k(x)$ by changing the signs of the $\ve_k$ in
(\ref{k}), so that
\[{k(x)\ov j(x)}=\prod_{k=0}^{m-1}\Ph(x,x_k)^{\ve_k}.\]
Recalling the normalizations (\ref{normalizations}) we deduce that
\[\im(k_0-j_0)=\sum_{k=0}^{m-1}\ve_k\,\lim_{x\ra\iy+0i}x\,\tl{g}(x,\,x_k),\]
and so
\bq\re\cI_1+|J|=2\pi a_2s+\pi\sum_{k=0}^{m-1}\ve_k\,\lim_{x\ra\iy+0i}
x\,\tl{g}(x,\,x_k).\label{cI1form}\eq\sp

\noi{\em Proof that} $\cI_2=O(s^{-1})$:\sp

Since $G_0(\bar{x})=-\bar{G_0(x)}$ and $G(x)=G_0(x)-a_1$ we have, with an
obvious notation,
\[\bar{h_+(x)}\,h_-(x)=\bar{k_+(x)}\,k_-(x)\,e^{2is(G_+(x)+a_1)}.\]
Again we use the facts that the product of the exponential Green
and Neumann functions in (\ref{k}) is smooth, uniformly in $s$, and that the
local
behavior of the other functions arising here is determined
locally. In particular $\bar{k_+(x)}\,k_-(x)$ is $O(|x|^{-1})$ at infinity, in
fact
for each $n$ its $n'th$ derivative is $O(|x|^{-n})$, and it has the behavior of
$(x-\al_k)^{-\hf}$ near $\al_k$ and analogous behavior near $\be_k$. The
function
$G_+(x)$ has the behavior of $x$ near infinity by (\ref{Gasymp1}), and
\[G'(x)={p(x)\ov\sqrt{q(x)}},\]
which is positive on $E^+$. It follows easily from these facts that if $\ph(x)$
is
any $C^{\iy}$ function with compact support which is identically 1 on a
neghborhood
of $\pl J$ then
\[\int\,(1-\ph(x))\,\bar{k_+(x)}\,k_-(x)\,e^{2isG_+(x)}\,dx=O(s^{-n})\]
for any $n$, while
\[\int\,\ph(x)\,\bar{k_+(x)}\,k_-(x)\,e^{2isG_+(x)}\,dx\]
has the behavior of
\[\int_0^{\iy}\ps(x){e^{-s\sqrt{x}}\ov\sqrt{x}}\,dx,\]
where $\ps$ is a $C^{\iy}$ function with compact support. Such an integral is
$O(s^{-1})$.\sp

\noi{\em Recapitulation}:\sp

We have shown that $\cI$, the integral in (\ref{hint}), equals
\[2\pi a_2s+\pi\,\sum_{k=0}^{m-1}\ve_k\,\lim_{x\ra\iy+0i}x\,\tl{g}(x,\,x_k)
-|J|+O(s^{-1}).\]
In particular $\cI=O(s)$ and so from Lemma 7 we get
\[\int_E|R(x)-e^{isx}|^2\,dx=\cI+O(1),\]
and this and (\ref{detform}) give
\[-\pi{d\ov ds}\log\,\det\,(I-K_s)=\cI+|J|+O(1).\]
If the error term $O(1)$ here were $O(s^{-1})$ we would have established
(\ref{Det})
with
\bq c_1=2\,a_2,\quad c_2(s)=\sum_{k=0}^{m-1}\ve_k\,\lim_{x\ra\iy+0i}
x\,\tl{g}(x,\,x_k).\label{c2}\eq
As it stands, of course, we have only proved the weaker statement
\[-{d\ov ds}\log\,\det\,(I-K_s)=c_1s+O(1).\]

\noi{\em Conclusion}:\sp

Here is how to get the stronger statement. The details of the analogous
argument
for orthogonal polynomials can be found in \cite{w1}.

It follows from (\ref{Rform}) that
\[|\int_Jg(x)\,e^{isx}\,dx|\leq\norm g\norm\,\norm R(x)-e^{-isx}\norm\]
for all $g\in\Es$, the norms being that of the space $L_2(E)$, and that this
becomes
an equality when $g(x)=R(x)-e^{-isx}$. From
this and (\ref{detform}) we deduce
\[-\pi{d\ov
ds}\log\,\det\,(I-K_s)=|J|+\max_{g\in\Es}{|\int_Jg(x)\,e^{isx}\,dx|^2\ov
\norm g\norm^2}.\]
Denote the maximum on the right by $M_0$. It follows from Lemma 4, in the form
\[\int_E\bar{(h_+(x)+h_-(x)-e^{-isx})}\,g(x)\,dx=\int_Jg(x)\,e^{isx}\,dx\]
(first for Fourier transforms of smooth functions in $L_2(-s,\,s)$ and then by
continuity for all such functions), that
\[ M_0\leq\int_E|h_+(x)+h_-(x)-e^{-isx}|^2\,dx.\]
The right side is just what we have called $\cI$ and so we know that
\bq M_0\leq\pi c_1+\pi c_2(s)-|J|+O(s^{-1}).\label{M0}\eq

Now we modify the the extremal problem by replacing $\norm g\norm$ by the norm
of
$g$ in $L_2(E,w_{\dl})$ where the weight function
$w_{\dl}(x)$ is equal to 1 except on the $2\dl$-neighborhood of $\pl E$, equal
to
$q(x)^{-\hf}$ in the $\dl$-neighborhood of $\pl E$ and, say, linear in between.
Denote by $M_{\dl}$ the corresponding extremum for the weight function
$w_{\dl}$. Clearly
\bq M_{\dl}\leq M_0.\label{M}\eq

There exist reproducing functions analogous to $k_\Ga$ in (\ref{k}) for any
weight
function satisfying quite general conditions, so that $|dx|$ in the statement
of
Lemma 3 can be replaced by $w_{\dl}(x)\,|dx|$. The extra ingredient is the
introduction
of a certain nonzero function in $\cH$ whose absolute value on $E$ equals
$w_{\dl}$.
We then proceed exactly as before, defining the analogous $h_s$ by (\ref{hs})
and $q$ by (\ref{q}). The main point is that because $w_{\dl}$ has the behavior
of $q(x)^{-\hf}$ near $\pl E$ the corresponding reproducing function $k_{\Ga}$
has
the behavior of $q(x)^{\hf}$, with the result that the factor $t^{1/4}$ does
not
appear in the denominator in (\ref{est}). This improves the error estimate in
Lemma 6 to $O(s^{-2})$ in this case, and so in the end we obtain
\[ M_{\dl}=\pi c_1+\pi c_{2,\,\dl}(s)-|J|+O(s^{-1}),\]
where $c_{2,\,\dl}(s)$ is the function in (\ref{c2}) associated with the weight
function $w_{\dl}$. Because of the continuity of this function in the class
$\Ga$
as well as the weight function $w$, we deduce that for any $\ve>0$ there is a
$\dl$
such that
\[M_{\dl}\geq\pi c_1+\pi c_2(s)-|J|-\ve+o(1).\]
Putting this together with (\ref{M0}) and (\ref{M}) gives the desired result.

When $J$ is a single interval of length 1, say $J=[-\hf,\,\hf]$, then $G(x)=
\sqrt{x^2-{1\ov4}}$, and we find that $c_1=1/4,\ c_2=0$. Thus (\ref{Det}) in
this case
is (\ref{det}) with the improved error term $O(s^{-1})$. Notice that this
is consistent with (\ref{dy}) since the first power of $s$ does not
appear in the exponent.

It is easy to see that $c_2(s)$ is periodic when $m=2$ and to compute its
period.
For its value depends only on the class of $e^{isG(x)}$, which in turn depends
only on
the value modulo $2\pi$ of $s$ times the quantity $\De_1G_0$ given by
(\ref{periods}).
Hence the period of $c_2(s)$ equals
\bq\left\{{1\ov\pi}\int_{\be_1}^{\al_2}{p(x)\ov\sqrt{q(x)}}\,dx\right\}^{-1}.
\label{period}\eq\sp

\begin{center}{\bf V. Demonstration of (\ref{Tr})} \end{center}\sp

It follows from Lemma 9(b) that for each $\ve>0$ the contribution to the
integral in
(\ref{trform}) of a sufficiently small neighborhood of $\pl J$ is $O(e^{\ve
s})$.
The main contribution will be from certain interior points of $J$, as we shall
see.
Outside any neighborhood of $\pl J$ the asymptotics of $R(x)$ are given by
Lemma 9(a),
which may be rewritten as
\[R(x)=e^{-isG(x)}\,(k(x)+O(s^{-\hf})).\]
It follows from general considerations that the derivative of the $O$ term
above is
also $O(s^{-\hf})$. We use the formula
\[(uv)'\,\bar{uv}=|u|^2\,({u'\ov u}|v|^2+v'\bar{v})\]
with $u$ equal to the exponential factor above and $v$ the other to write
\[R'(x)\,\bar{R(x)}=e^{2s\,\im G(x)}\left\{-isG'(x)\,|k(x)+O(s^{-\hf})|^2
+k'(x)\,\bar{k(x)}+O(s^{-\hf})\right\}.\]
Since $G'(x)$ is purely imaginary in $J$ the first term in the braces is purely
real,
and so does not contribute to the right side of (\ref{trform}),
and we are left with the computation of
\bq\int e^{2s\,\im G(x)}\,(\im
k'(x)\,\bar{k(x)}+O(s^{-\hf}))\,dx.\label{trint}\eq
The integration is taken over $J$ with its little neighborhood of $\pl J$
removed. The
main contribution to the integral will come from the point or points where $\im
G(x)$
achieves its maximum, and these will be among the zeros of $p(x)$. Denote these
zeros
by $z_1,\cdots,z_m$, one lying in each interval of $J$.  Since at each
$z_i$
\bq{d^2\ov dx^2}\im G(x)=\im {d\ov
dx}{p(x)\ov\sqrt{q(x)}}=-{|p'(x)|\ov\sqrt{|q(x)|}},
\label{ders}\eq
we find by standard asymptotics that (\ref{trint}) equals
\[\sqrt{{\pi\ov s}}\sum_{i=1}^me^{2s\,\im G(z_i)}\left\{{|q(z_i)|^{1/4}\ov
|p'(z_i)|^{\hf}}\im k'(z_i)\,\bar{k(z_i)}+O(s^{-\hf})\right\}.\]
We have from (\ref{k})
\[|k(z_i)|^2=\hf\prod_{k=1}^{m-1}|\Ps(z_i,\,x_k\st\,x_k)\,\Ph(z_i,\,x_k)^{\ve_k}|,\]
and, using in addition to (\ref{k}) the very last part of (\ref{ders}), we find
that
\[\im {k'(z_i)\ov k(z_i)}=\hf\left\{-{|p'(z_i)|\ov\sqrt{|q(z_i)|}}+
\sum_{k=1}^{m-1}(\tl{N}(z_i,\,x_k\st\,x_k)+\ve_k\,\tl{g}(z_i,\,x_k\st\,x_k))
\right\}.\]
Combining these last two formulas gives $\im k'(z_i)\,\bar{k(z_i)}$. Comparing
with (\ref{trform}) shows that (\ref{Tr}) is established, with
\[c_3=2\;\mbox{max}_k\ \im G(z_k)\]
and
\[c_4(s)={1\ov4\sqrt{\pi}}\sum_{i}\left\{{|q(z_i)|^{1/4}\ov|p'(z_i)|^{\hf}}
\left[{|p'(z_i)|\ov\sqrt{|q(z_i)|}}-\sum_{k=1}^{m-1}(\tl{N}(z_i,\,x_k\st\,x_k)+
\ve_k\,\tl{g}(z_i,\,x_k\st\,x_k))\right]\right.\]\bq\left.\times
\prod_{k=1}^{m-1}|\Ps(z_i,\,x_k\st\,x_k)\,\Ph(z_i,\,x_k)^{\ve_k}|\right\},\label{c4}\eq
where the outer sum is taken over those $i$ for which
\[\im G(z_i)=\mbox{max}_k\,\im G(z_k).\]

There is an explicit representation of the quantities $\im G(z_k)$. It follows
from
(\ref{conds}) and (\ref{G}) that for $z\in(\al_k,\,\be_k)$
\[\im G(z)=\int_{\al_k}^z{\pm p(x)\ov\sqrt{|q(x)|}}\,dx,\]
the sign being that of $p(\al_k)$. The maximum of this, its value at $z_k$, is
the
integral to $z_k$, while
the integral from $z_k$ to $\be_k$ is the negative of this since the sum of the
two
is zero. Hence
\[\im G(z_k)=\hf\int_{\al_k}^{\be_k}{|p(x)|\ov\sqrt{|q(x)|}}\,dx,\]
and so
\[c_3=\mbox{max}_k\ \int_{\al_k}^{\be_k}{|p(x)|\ov\sqrt{|q(x)|}}\,dx.\]
When $J$ is a single interval of length 1 this equals 1.

When $m=1$ (\ref{c4}) becomes
\[{1\ov4\sqrt{\pi}}{|p'(z)|^{\hf}\ov|q(z)|^{1/4}},\]
simply, and when the length of $J$ is 1 this is $1/2\sqrt{2\pi s}$ and we
recover
the formula (\ref{tr}). When $m=2$ the function $c_4(s)$ is periodic with the
same
period (\ref{period}) as $c_2(s)$.

\end{document}